# A comprehensive study of the (2√3x2√3)R30° structure of silicene on Ag(111)

**H Jamgotchian[1], B Ealet[1], Y Colignon[2], H Maradj[1,3], J-Y Hoarau[1], J-P Biberian[1] and B. Aufray[1]**

[1] Aix-Marseille Université, CNRS, CINaM, UMR 7325, 13288 Marseille, France
[2] Aix-Marseille Université, CNRS, IM2NP, UMR 7334, 13397 Marseille, France
[3] LSMC, Université d'Oran es-sénia, 31100, Oran, Algeria

E-mail: jamgotchian@cinam.univ-mrs.fr

**Abstract.** The deposition of one silicon monolayer on Ag(111) gives rise to a set of superstructures depending on growth conditions. These superstructures are correlated to the epitaxy between the honeycomb structure of silicon (so called silicene) and the silver substrate. In this paper, from a detailed re-analysis of experimental results, obtained by Scanning Tunneling Microscopy and by Low Energy Electron Diffraction on the (2√3x2√3)R30° structure, we propose a new atomic model of the silicene layer based on periodic arrangements of perfect areas of (2√3x2√3)R30° surrounded by defect areas. A generalization of this model explains the main experimental observations: deviation of the average direction, Moiré patterns and apparent global disorder. In the frame of the proposed model, the apparent disorders observed on the STM images, would be topological effects, i.e. the silicene would keep a quasi-perfect honeycomb structure.

## 1. Introduction

The silicene, the counterpart of graphene for silicon is mainly obtained by epitaxial growth of silicon on oriented metallic substrates [1,2]. Today, silver is still the most widely used substrate even though silicene has been also synthetized with success on Ir(111) [3], $ZrB_2(111)$ [4] and $MoS_2$ [5]. Silver is an appropriate substrate due to its strong tendency to undergo phase separation (no silicide formation) and to the quasi-perfect match between four silver inter-atomic distances (1.156 nm), and three silicon inter-hexagons distances (1.152 nm). On a silver substrate, the first observations of

"graphite-like honeycomb structure" were obtained on Ag(100) (nano-stripes) [6], then on Ag(110) (nano-ribbons) [7] and finally on Ag(111) (silicene sheet) which is today the most studied orientation [8–19].

With the (111) orientation, four structures are generally observed depending on growth temperature and deposition rate. We have shown that all these structures can be explained by different orientations of the silicene layer with respect to the silver substrate [12]. These orientations of the silicene layer are characterized by the rotation angle (α) between the Si[110] and the Ag[110] directions. Starting from a (4x4) structure, a rotation of the silicene layer relative to the silver substrate gives rise successively to the following structures: a (√13x√13)R13.9°, a (2√3x2√3)R30°, a (√7x√7)R19.1° (which is rarely observed) [20] and finally a second (√13x√13)R13.9°.

**Table 1.** Parameters of different unit cell structures[a]

| Structures relative to Ag | α (°) | $L_{Ag}$[b] (nm) | $L_{Si}$[c] (nm) | $L_{Si}/L_{Ag}$[d] |
|---|---|---|---|---|
| (4x4) | 0 | 1.156 | 1.152 | 0.997 |
| (2√3x2√3)R30° | -10.9 | 1.001 | 1.016 | 1.015 |
| (2√3x2√3)R30° | +10.9 | 1.001 | 1.016 | 1.015 |
| (√13x√13)R+13.9° Type I | -27 | 1.042 | 1.016 | 0.975 |
| (√13x√13)R+13.9° Type II | -5.2 | 1.042 | 1.016 | 0.975 |
| (√13x√13)R-13.9° Type I | +27 | 1.042 | 1.016 | 0.975 |
| (√13x√13)R-13.9° Type II | +5.2 | 1.042 | 1.016 | 0.975 |

[a] Bulk Si ($d_{HSi}$ = 0.384 nm) and bulk Ag ($d_{Ag}$ = 0.289 nm) distances are used as references.
[b] Length of the unit cells based on the silver structure.
[c] Length of the unit cells based on the bulk silicon structure.
[d] Strain factor.

In order to know if the silicene layer is compressed or expanded, we have calculated the strain factor $L_{Si}/L_{Ag}$ as shown in table 1. The ratio shows that the silicene layer should be expanded for the (4x4) and (√13x√13)R13.9°, and contracted for the (2√3x2√3)R30° structures. The formation and the existence of all these structures are strongly dependent on substrate temperature [12]. The domain of existence of quasi-pure (4x4) is between ~180°C and ~220°C. In the 220°C-280°C range, there is a mix of (4x4), (√13x√13)R13.9° and (2√3x2√3)R30° structures. Above 280°C a quasi-pure (2√3x2√3)R30° is obtained. In this temperature domain, some studies have shown perfect LEED patterns of (2√3x2√3)R30° with large spots [12,17]. Others have shown sharp LEED patterns of the (2√3x2√3)R30° with some additional spots which have been interpreted as a mixture of (√19x√19)R23.4° and (3.5x3.5)R26° structures [16,17]. Recently the LEED pattern with extra spots has been interpreted as (√133x√133)R4.3° and the formation of a surface alloy [21]. However, under the same conditions, low resolution STM images show apparent ordered (2√3x2√3)R30° structures but with an average angle which can be different from the expected 30° angle [22] and, sometimes, associated with a Moiré pattern [23]. On high resolution STM images, the structures appear locally ordered with the expected morphology [9], but surrounded by many defects [22]. One paper claims that this structure is not ordered enough to exist and then "*die suddenly*" [24]. However, a more recent study has shown that this structure can be used to build a Field Effect Transistor [25].

The present paper is focused on the (2√3x2√3)R30° structures obtained after deposition of silicon with substrate temperatures at 390°C, 370°C and 270°C. From a detailed analysis of STM images (low and high resolutions) and from LEED observations, we propose in this paper a comprehensive atomic model of the silicene layer forming the (2√3x2√3)R30° structure. This atomic model is based on a periodic arrangement of perfect areas of (2√3x2√3)R30° surrounded by defect areas. A generalization of this model explains remarkably well the experimental observations: deviations of the average directions of the structure, Moiré patterns, and apparent global disorders.

## 2. Experimental set-up

The experiments were performed in an ultra-high vacuum (UHV) system equipped with Low Energy Diffraction (LEED), Auger Electron Spectroscopy (AES) and Room Temperature Scanning Tunneling Microscopy (Omicron RT-STM 1). The protocol used for the Ag(111) surface preparation is the usual one [12]. Silicon deposition is performed by thermal evaporation from a silicon wafer heated by the Joule effect. The monolayer (ML) coverage is checked by AES. The sample is heated by a tungsten resistor located under the sample holder. Because the observed structures are very sensitive to the substrate temperature and since the thermocouple is located close to the sample holder, the temperature has been recalibrated using the melting temperature of two metals (Zn and Pb), positioned in place of the Ag sample. The possible drift of the STM images is corrected using the hexagonal symmetry of the surface checked by FFT. Such a correction increases the precision of the angle measurements.

## *3.* **Experimental results**

The first silicon deposition has been performed with a substrate temperature at 390°C. The low resolution STM image is shown in figure 1(a) and the corresponding LEED pattern in figure 1(b). The FFT of the STM image is shown on figure 1(c). On the STM image, the Ag[110] direction is identified by a yellow line. The average direction of the structure (red line) is at 25.5 ± 1° with respect to the Ag[110] direction. This is different from the 30° expected from a perfect (2√3x2√3)R30° structure direction (white line). On this image, bright and dark areas create a Moiré pattern. Bright areas, which correspond to perfect (2√3x2√3)R30° areas, are separated by defect areas (darker areas). A pink diamond (3.7 ± 0.1nm) highlights the unit cell of the Moiré pattern. The LEED pattern (figure 1(b)) shows sharp spots characteristic of a well-ordered structure. Some doublets and triplets (inside white circles) close to the expected spots of a perfect (2√3x2√3)R30° structure are visible. Since those doublets and triplets are also visible on the FFT of the STM image of figure 1(c), the LEED pattern is indeed a signature of the Moiré. The white line (figure 1(c)) crosses the middle of the triplets and the red line is exactly on one of a triplet spots. The discussion section presents a detailed analysis of the LEED.

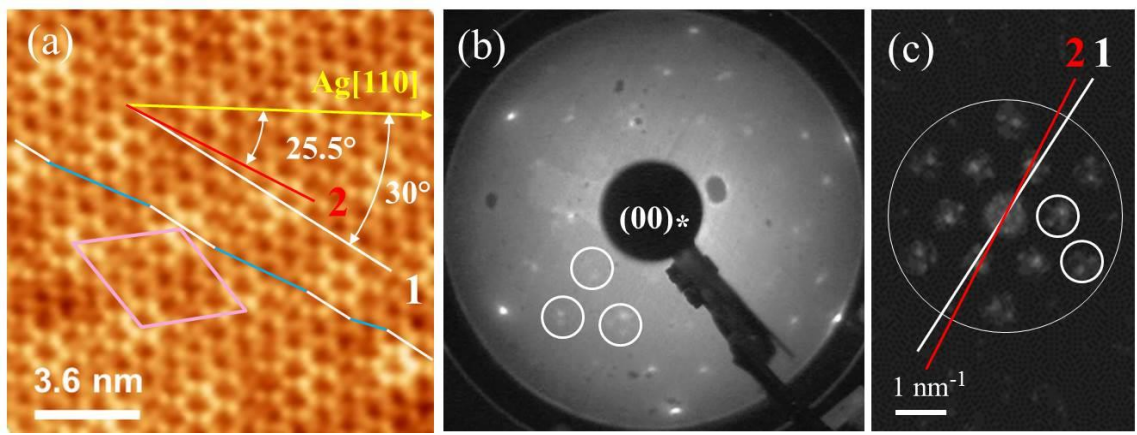

**Figure 1.** (2√3x2√3)R30° structure obtained after deposition of one monolayer of silicon on Ag(111) at $T_{Sub}$ = 390°C. (a) Typical large view STM image of the structure showing a Moiré pattern (V=-1.4V; I=0.1nA). Pink diamond is the unit cell of the Moiré pattern. White-blue zigzag line connects the different domains (perfect and defect areas). White lines (# 1) are the direction of a perfect (2√3x2√3)R30° structure. Red lines (#2) correspond to the average direction of the observed structure. (b) Experimental LEED pattern (E = 78 eV). (c) FFT of the STM image shown in (a).

Figure 2(a) shows a high resolution STM image after deposition of one monolayer of silicon at a slightly lower temperature of 370°C. Figure 2(b) is a blow-up of the blue rectangle of figure 2(a). The corresponding LEED and FFT patterns are presented in figure 2(c) and 2(d) respectively. The LEED pattern shows a quasi-pure (2√3x2√3)R30° structure with large and fuzzy spots indicating that the ordered domains are non-uniform in size. In the STM image, we identify locally perfect domains of (2√3x2√3)R30° structures highlighted in yellow. Within each domain, the direction of the structure is exactly 30° (white line) and the distance between two black holes is in perfect agreement with the expected value of the (2√3x2√3)R30°structure (1.00 ± 0.05 nm). Each perfect area is surrounded by defect zones giving rise to a superstructure which unit cell is highlighted by a pink diamond (3.8±0.1 nm). Note that without highlighting the perfect areas on the STM image, such a periodicity would not be easy to observe. The shift between two next nearest neighbors of perfect (2√3x2√3)R30° areas is ≈0.4 nm, i.e. 1.5 interatomic silver distance (0.434 nm). The red line indicates the average direction of the structure (34.5 ± 1°). The FFT of the STM image of figure 2(a) is shown on figure 2(d). The white circles evidence the extra spots similar to those observed in the figure 1(c). Here again, the white line (figure 2(d)) crosses the triplets and the red line goes exactly over one of the triplet spots.

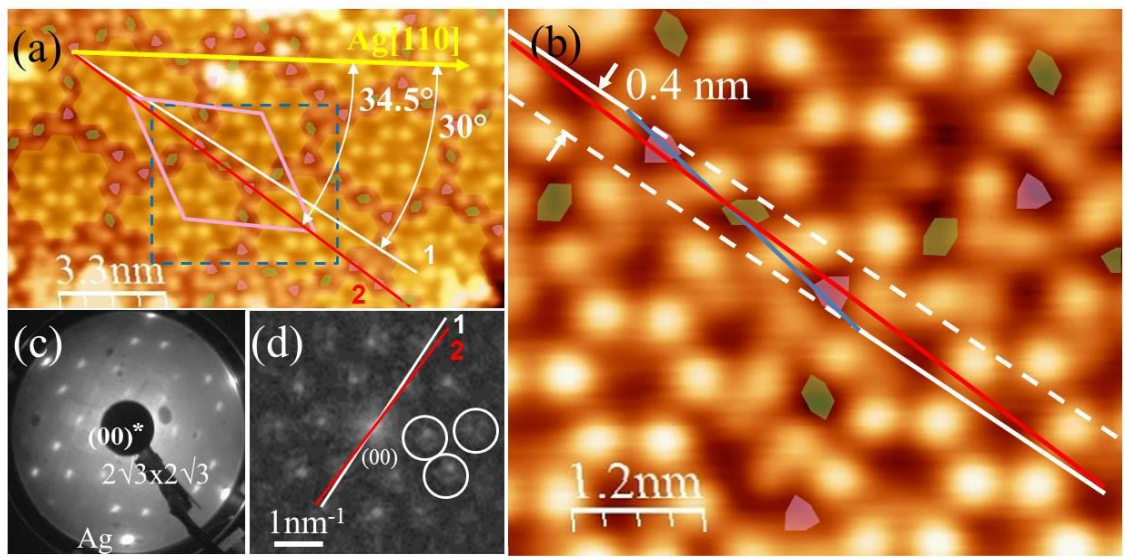

**Figure 2.** (2√3x2√3)R30° structure obtained after deposition of one monolayer of silicon on Ag(111) at $T_{Sub}$ = 370°C. (a) High resolution STM image (V = -1.7V, I = 1.1nA) showing the different defects areas: linear defects in green and point defects in pink. White lines (# 1) are the direction of a perfect (2√3x2√3)R30° structure. Red lines (#2) correspond to the average direction of the observed structure. (b) Blow-up of the blue rectangle of (a). (c) Corresponding LEED pattern (E = 58eV). (d) FFT of the STM image shown in (a).

Figure 3(a) shows a low resolution STM image obtained after deposition of a Si monolayer at 280°C. Such an image is not representative of the whole surface which contains different structures as corroborated by the LEED (figure 3(b)) showing a mix of (2√3x2√3)R30°, (√13x√13)R13.9° and (4x4) structures. Similarly to figure 1(a) and figure 2(a), the yellow line corresponds to the [110] direction of silver and the white lines to the direction of the perfect (2√3x2√3)R30° structure. The average orientation (22±1°) of the structure (red line) is far from the expected directions of either the (√13x√13)R13.9° or the (2√3x2√3)R30° structures (blue and white lines respectively). From this type of image, it is not possible to identify the structure. Furthermore, the periodicity i.e. the distance between two black holes, is smaller (0.96 ± 0.05 nm) than the expected ones (1.001 nm and 1.042 nm for (2√3x2√3)R30° and (√13x√13)R13.9° respectively - table 1). Interestingly, the FFT of the STM image (figure 3(c)) shows extra spots (inside the white circle) evidencing the presence of a superstructure not directly seen on the STM image. On the FFT, the extra spots inside the white circles are similar to those observed on figure 1(a) and 2(a). This confirms that a similar superstructure exists related to a (2√3x2√3)R30° structure, but not visible on the STM images. From the FFT, we deduce a size of the unit cell of this superstructure (≈1.8 nm).

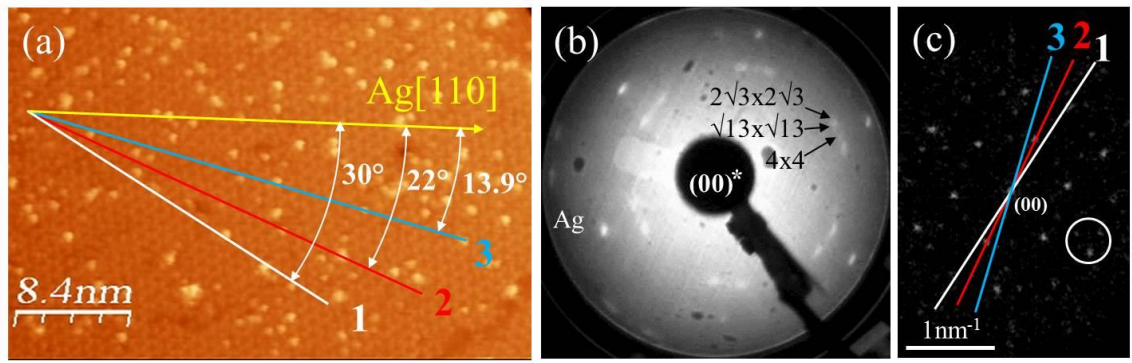

**Figure 3.** (2√3x2√3)R30° structure obtained after deposition of one monolayer of silicon on Ag(111) at $T_{Sub}$ = 280°C. (a) Large view STM image (V = -1.1V, I = 0.8nA). White lines (# 1) are the direction of a perfect (2√3x2√3)R30° structure. Red lines (#2) correspond to the average direction of the observed structure. Blue lines (# 3) correspond to the direction of a perfect (√13x√13)R13.9° structure. (b) Corresponding LEED pattern (E = 58eV). (c) FFT of the STM image shown in (a).

## 4. Periodic local relaxation model

These three experiments show that a small variation of growth temperature has a large effect on the observed superstructures of the silicene layer characterized by LEED and STM. In this section, we show that all these structures are constituted of perfect (2√3x2√3)R30° domains separated by defect areas. Only the average size and the size distribution of the ordered domains depend on growth temperature. We propose an atomic model, which explains in a coherent manner our experimental observations as well as those of other experimentalists.

*4.1. Presentation of the model*

The perfect (2√3x2√3)R30° structure presents two domains (α = ±10.9°). These two domains exist on the silver surface but cannot be distinguished by STM since only the top silicon atoms are visible [12,18,23,26]. Nevertheless, as shown above in the previous section, on the observed structures, the angles are different from the expected ones. In the following, we detail the defects at the atomic scale, interpreting them by local deformation/relaxation of the silicene layer.

The geometrical atomic model proposed to explain the above results is based on the fact that on silver substrates, only the silicon atoms on top sites are visible in STM images and on the hypotheses that the silicene layer keeps a honeycomb structure. In addition, we assume that the silver substrate remains rigid under the silicene layer (no deformation). To build up the model, we start from the high resolution STM image of the figure 2(a). The large regular hexagons are perfect (2√3x2√3)R30° areas, in agreement with theoretical studies [9,26–28]. Each of them is constituted of seven silicene hexagons.

In defect areas, the large hexagons, which are highlighted in green and in pink in figures 2(a,b), appear shrunken and deformed. They have the appearance of a contraction of the structure. Each of them is interpreted as a local and small deformation of the silicene hexagons. However, because in a perfect (2√3x2√3)R30° structure the silicene is contracted by 1.6% compared to bulk Si, we attribute the local relaxation of the silicene layer to a dilatation of the layer. To understand this local relaxation mechanism, we analyze the deformation in two steps in the case of α = -10.9°. In the first step, we consider a linear deformation in one direction. Figure 4 presents the comparison between a perfect (2√3x2√3)R30° structure (figure 4(a)) and the same area in which the central part (green area) has undergone a deformation, obtained by a continuous dilatation of the silicene layer in the armchair direction towards the top of the figure (blue arrow). This results in the appearance of two perfect areas (lower and upper parts) and a defect area highlighted in green, in which the dashed silicon atoms indicate the previous atomic positions before dilatation towards the new positions in orange. The resulting top atoms form a shrunken hexagon in agreement with the STM image of figure 2(b). This confirms that the apparent contraction of those hexagons corresponds to a dilatation of the silicene layer. In the upper part of figure 4(b), the large hexagons are shifted by one silver interatomic distance, which corresponds to a shift of the silicene layer by 1/3 of silver interatomic distance. In the defect area, for each Si atom, the mean displacement with respect to a perfect silicene layer (evidenced by dashed Si atoms) is only, in average ~ 0.03 nm which corresponds to an uniaxial local dilatation of ~ 14% of the silicene layer in armchair direction.

It is interesting to note that, in the linear defects, the distance between silicon top atoms a-b and c-d is $\sqrt{13}d_{Ag}$.

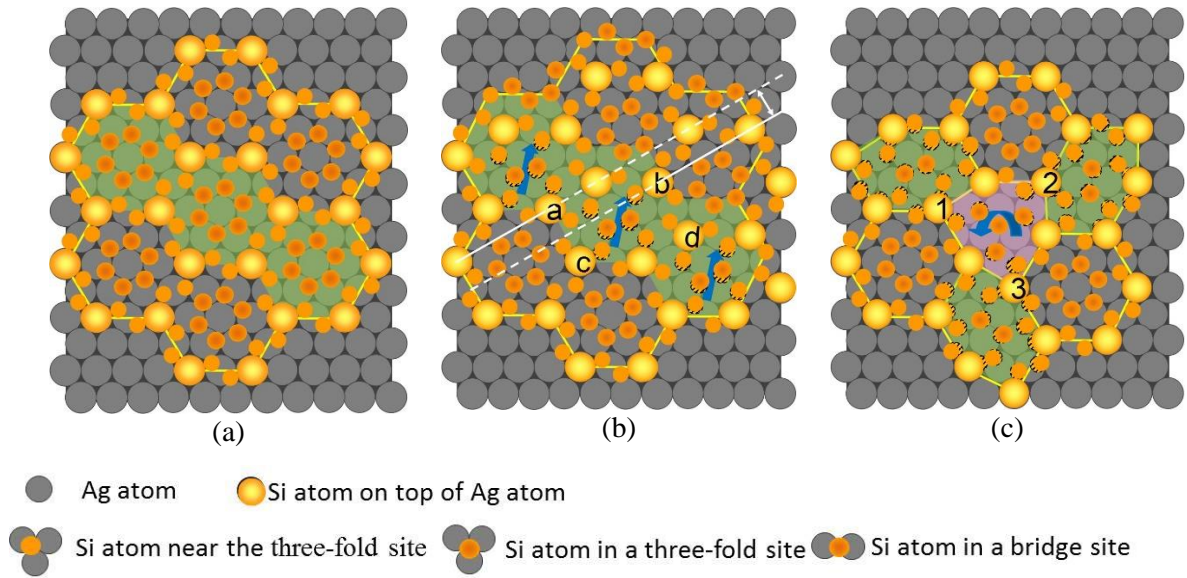

(a)   (b)   (c)

- Ag atom
- Si atom on top of Ag atom
- Si atom near the three-fold site
- Si atom in a three-fold site
- Si atom in a bridge site

**Figure 4.** Schematic presentation of a possible formation of local defects in the $(2\sqrt{3}x2\sqrt{3})R30°$ structure for $\alpha$ = -10.9°. (a) Perfect $(2\sqrt{3}x2\sqrt{3})R30°$ structure. The green area highlights where the relaxation will take place. (b) Linear defects obtained by dilatation in one direction (blue arrow) of a perfect silicene layer ($\approx d_{Ag}/3$). The atoms visible by STM (brighter and larger) appear as shrunken hexagons. (c) Point defects (pink) obtained by the combination of three linear defects (green) which correspond to a small local rotation of the silicene layer around a central silicon atom. The atoms visible by STM form a deformed hexagon (pink).

Due to the symmetry of the $(2\sqrt{3}x2\sqrt{3})R30°$ unit cell, the deformation exists necessarily in six equivalent directions (60°) which leads to a topological point defect of the structure as shown in figure 4(c). In this area, where silicon atoms are not visible by STM, there is a small rotation (anticlockwise for $\alpha$ <0, clockwise for $\alpha$ >0) of the hexagons around a central silicon atom located in a three-fold site. The corresponding STM images generated by the silicon atoms on top sites will produce a deformed hexagon highlighted in pink. Such defects are visible on the STM image of figure 2(b). Note again, that within the point defect, the distance between silicon top atoms 1, 2 and 3 is $\sqrt{13}d_{Ag}$. Due to the hexagonal symmetry of the perfect structure, there are three equivalent linear defects and two equivalent point defects, which surround the perfect areas. Equivalent linear defects are rotated by 120° and equivalent point defects are rotated by 180°.

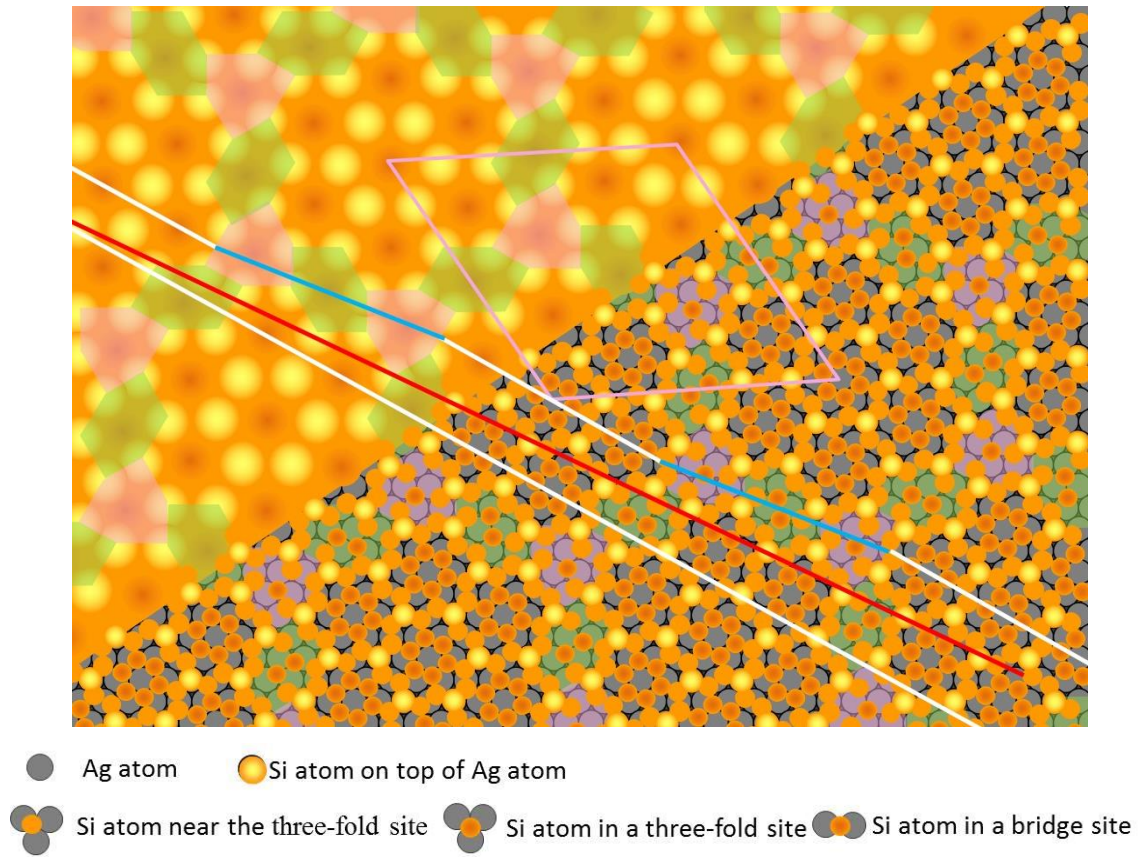

**Figure 5.** Large scale ball model of the $(2\sqrt{3}\times2\sqrt{3})R30°$ structure ($\alpha = +10.9°$) with perfect periodic areas surrounded by defects areas. The upper left part shows the expected STM image. The white-blue zigzag line connects the different domains (perfect and defect areas). The white line corresponds to the orientation of a perfect $(2\sqrt{3}\times2\sqrt{3})R30°$ structure and the red line to the average angle which should be observed on STM images at large scale. The unit cell pattern (pink diamond) is a $(\sqrt{133}\times\sqrt{133})R4.3°$ superstructure.

The application of the model in all directions induces a new superstructure. The example for $\alpha > 0$ shown in figure 5, corresponds to a ball model with seven perfect hexagons surrounded by periodic defect areas. The periodicity of the perfect and defect domains gives rise to a large $(\sqrt{133}\times\sqrt{133})R4.3°$ superstructure evidenced by the pink diamond in figure 5. Such a model is just geometrical and not a theoretical calculation.

In figure 5, the lower right part presents the atomic ball model and the upper left part represents the structure, which should be observed with STM at low resolution. The bicolor white-blue zigzag line connects the different domains (perfect and defect areas). The white line corresponds to the orientation of a perfect (2√3x2√3)R30° structure and the red line to the average angle which should be determined on corresponding STM images. The shift between two next nearest neighbors (2√3x2√3)R30° areas is 1.5 interatomic silver distance (0.434 nm) which is in good agreement with the measured distance between the two white lines on the STM image of figure 2(a) (≈0.4 nm).

Note that the atomic model proposed does not predict that the defect areas must appear darker than the perfect areas as observed on the STM image of figure 1. This suggests that the silicon atoms in the defect areas are not located exactly on top of silver atoms but close to them, inducing darker areas.

A dynamic analysis of the LEED patterns for such large superstructures being almost impossible to calculate, we used a simple kinematic approach, equivalent to a Patterson function [29,30], which is a convolution of the (√133x√133)R4.3° large superstructure unit cell with the perfect small local (2√3x2√3)R30° structure. As a consequence, on the LEED patterns, only the spots of the large superstructure close to the (2√3x2√3)R30° spots will be visible [31]. In fact both, the (2√3x2√3)R30° and the (√133x√133)R4.3° structures are commensurable with Ag(111), but not commensurable together.

Figure 6(a) is the expected LEED pattern for one domain of the (√133x√133)R4.3° superstructure ($\alpha$ = +10.9°). The yellow dots correspond to the silver substrate and the blue dots to the (√133x√133)R4.3° extra spots. Experimentally, the LEED pattern will be a combination of the two domains ($\alpha$ = ±10.9°), as shown in figure 6(b). Red dots evidence the agreement with the experimental LEED pattern of figure 1(b).

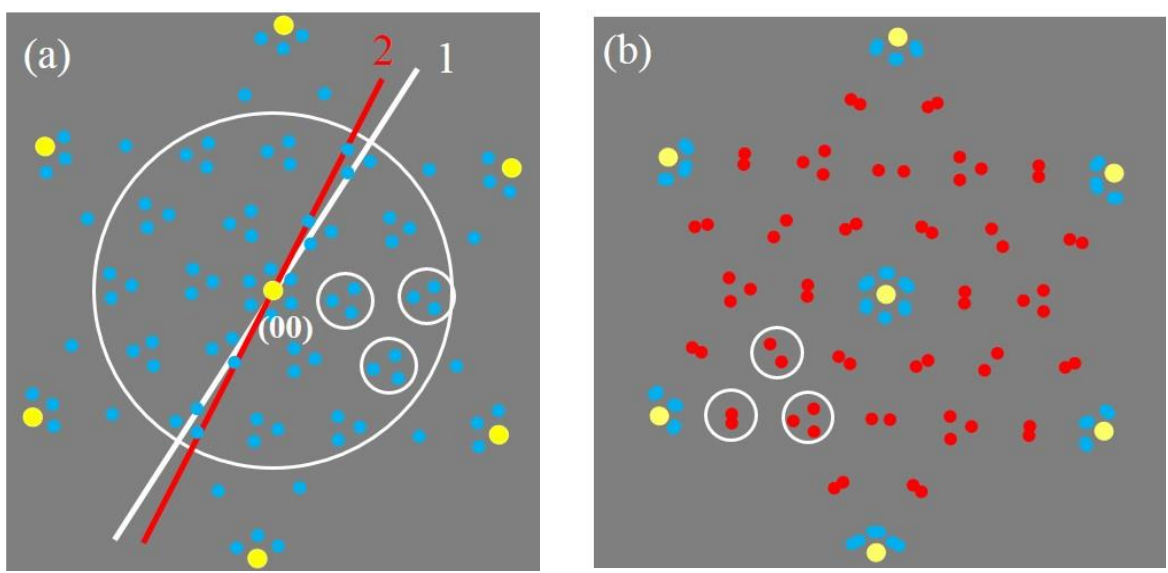

**Figure 6.** Expected LEED patterns of the (√133x√133)R4.3° superstructure (a) LEED pattern for α = 10.9°, where only the dots near the (2√3x2√3)R30° structure are shown (blue dots). Yellow dots belong to the Ag(111) substrate; a large white circle allows easy comparison with FFT of the STM images. The white and red lines are equivalent to those of figure 1(c). (b) LEED pattern for two domains, α = ±10.9°. Red dots correspond to the more intense spots visible in the experimental LEED patterns.

*4.2. Generalization of the model*

This model shows that the periodic deviation creates an average rotation of the structure which value depends directly on the size of the perfect areas. The sign of the angle of this average deviation (angle shift) and the sign of α are the same.

By changing the size of the perfect (2√3x2√3)R30° areas, the size of the new superstructure and the angle shift will change: the larger the perfect domains, the smaller the angle shift and the structure will go towards a perfect (2√3x2√3)R30° structure. In other words, the smaller the number of defects, the smaller will be the angle shift, which is always easier to measure on low-resolution STM images. On the contrary, for high-resolution STM images, it is possible to measure the quantity of defects but not the angle shift. Experimentally, the angle shift measurement allows estimating the average size of the perfect domains, which is equivalent to a statistical determination of the defects.

The model has been developed for the (√133x√133)R4.3° superstructure with hexagonal symmetry, constituted by seven large perfect hexagons of (2√3x2√3)R30° structure. The simplest superstructure with hexagonal symmetry is the (√31x√31)R8.9° superstructure constituted by one large perfect hexagon of (2√3x2√3)R30° structure. A generalization of the model allows us to predict the average angle (i.e. the angle shift) of the (2√3x2√3)R30° structure versus the distance between two perfect domains (i.e. the periodicity of the Moiré) for α < 0, and for α > 0 as it is shown on figure 7(a) (continuous lines). The (√31x√31)R8.9° and (√133x√133)R4.3° superstructures are indicated by vertical dashed lines. For an infinite number of hexagons a complete perfect (2√3x2√3)R30° structure would be formed. The sign of the angle shift is directly relative to the sign of α. When two domains with α < 0, and α > 0 are simultaneously observed on the same STM image, the angle between both domains will be equal to twice the angle shift as it has been shown in the figure 9 of [22].

Note that experimentally, for a given temperature, we observe a dispersion of the sizes of perfect areas as shown on figure 1(a) and figure 2(a). On this calculated curve, we have reported our experimental data and those issued from other studies (see discussion section). In the case of figure 3, where the Moiré pattern is not observable on the STM image, we have used the FFT, which reveals the periodicity. It is remarkable that this curve fits well all the experimental data.

With the same approach, figure 7(b) shows the expected variations of the average size of the structure (which is measured on low resolution STM images) as a function of the angle shift (continuous line). This curve shows that starting from a perfect (2√3x2√3)R30° structure the average distance should continuously decrease with the angle shift. On this calculated curve, we also report our experimental data and those from other studies (discussed in the following). Note again that the theoretical curve fits well our three experimental results and not very well the value issue from [24].

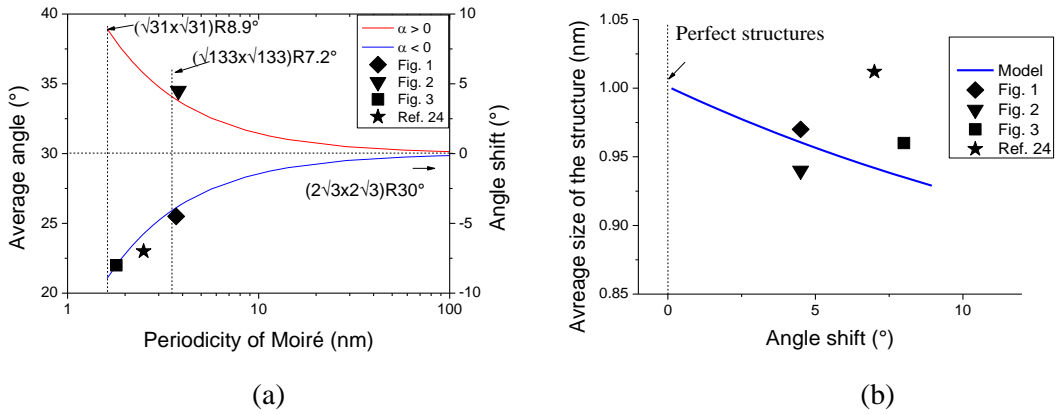

(a)            (b)

**Figure 7.** Calculated variations of the apparent (2√3x2√3)R30° structure parameters. (a) Average angle of the structure relative to Ag[110] (left axis) and angle shift (right axis) with respect to a perfect (2√3x2√3)R30° structure as a function of the Moiré periodicity (top curve $\alpha = +10.9°$, bottom curve $\alpha = -10.9°$). The size of the error bars is of the order of the symbols dimensions. (b) Variation of the average size of the (2√3x2√3)R30° structure as a function of the angle shift.

## 5. Discussion

The geometrical model that we propose allows a better understanding of the variety of (2√3x2√3)R30° structures observed by us and various groups [12,16,17,20,22–24]. Concerning our own experiments, for three different deposition temperatures, we interpret the STM images with the same model. However, the LEED patterns show extra spots only at the highest growth temperature. On the LEED pattern of figure 2(c) the extra spots are not visible due to the likely dispersion of the size of the perfect domains, as shown on the STM image (figure 2(a)). Probably for the same reason and the presence of the two other structures, (4x4) and (√13x√13)R13.9°, the extra spots are not visible on the LEED pattern of figure 3(b). Note that on the contrary, the FFT characteristic of well-defined local areas (figure 2(c) and figure 3(b)) are in good agreement with the model of figure 6(a). In the following, we analyze LEED and STM results of other groups.

A LEED pattern, showing extra spots close to the (2√3x2√3)R30° structure, has also been observed by Arafune et al. (figure 3(c) in [16]), Moras et al. [17] and recently by Rahman et al. [21]. Arafune et al. [16] and Moras et al. [17] interpret the LEED patterns as a mixture of various structures, including (√19x√19)R23.4° and (3.5x3.5)R26°. We propose that the diffraction spots of these LEED patterns are from (√133x√133)R4.3° (or close to) structures. Finally, Rahman et al. [21] have indexed a similar LEED pattern with a same large (√133x√133)R4.3° superstructure. Without STM observations, they suggest from Auger Electron Spectroscopy the formation of a surface alloy. Knowing that at high temperature deposition, there is dissolution of silicon in the silver substrate [24], the saturation of the Auger silicon signal is probably not the signature of a surface alloy formation but more likely due to a partial dissolution of silicon in silver.

An equivalent Moiré pattern is observed by Feng et al. (figure 1(b) in [23]). The measured Moiré periodicity (≈3.8 nm) is very close to ours (≈3.7 nm). They interpret this Moiré by a periodic succession of perfect and defect areas. Unfortunately, they do not show a complete high resolution STM image of the defect areas and, they do not mention the average deviation of the angle of the structure: "*the angle between the direction of Moiré pattern and honeycomb structure is about 30°*" [23]. Concerning the periodicity of the structure they mention "*lattice period about 1.0 nm are observed at the bright part of the Moiré pattern*" [23]. Therefore, we could not plot their values on the curve.

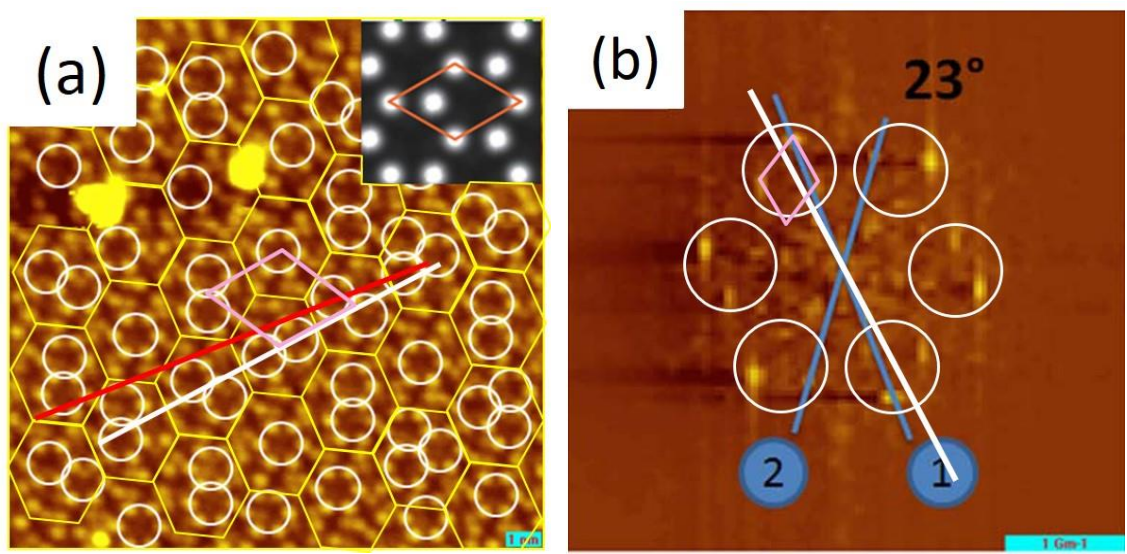

**Figure 8.** (2√3x2√3)R30° structure reproduced and modified from Liu et al. [24]. (a) We have superimposed on the STM image a hexagonal periodic lattice (yellow tram) showing a long distance periodicity given by pink diamond, as well as a red line corresponding to the average structure

direction (23°) given by the authors and a white line corresponding to the perfect (2√3x2√3)R30° structure direction. (b) FFT given by authors in which we add the white line and pink diamond.

Finally, Liu et al. [24] report the growth of a highly disordered (2√3x2√3)R30° structure. The authors interpret the STM image (figure 2 in [24]) as a mixture of (2√3x2√3)R30° and (√19x√19)R23.4° structures. Figure 8 is a copy of their STM image and its FFT. The FFT presents triplets, similar to the ones of our experiments (figure 1(c), figure 2(d) and figure 3(c)), that we surrounded by white circles. We have drawn a white line corresponding to the perfect (2√3x2√3)R30° direction. Using the model, if we assume that the triplets come from a Moiré pattern, the extra spots give its periodicity (~2.5 nm). When we superimpose a hexagonal lattice (yellow) with periodicity of 2.5 nm (pink diamond), all the perfect areas of (2√3x2√3)R30° structures (surrounded by white circles by authors) coincide with this lattice. This reveals a quasi-periodic large structure composed of small perfect areas of (2√3x2√3)R30° structure surrounded by defect areas as proposed in the model. From figure 7(a) we deduce an angle shift of 24°, in perfect agreement with the angle given by the authors of 23°. The apparent disordered structure obtained by Liu et al. [24] is not so surprising since they have used growth conditions at a temperature such that large Moiré structures cannot be formed.

In our model a linear defect corresponds to a uniaxial stain and a point defect corresponds to a biaxial one. Recent DFT calculations on similar homogeneous strains [32,33] show strong modifications of the electronic structure of standalone silicene layer.

Finally, Tao et al. [25] have built a field effect transistor with a silicene layer having a (2√3x2√3)R30° structure. Their STM image of the (2√3x2√3)R30° (figure 2(e) of [25]) shows many apparent defects close to the one shown in figure 8a. Nevertheless, Raman spectroscopy signatures of this structure are characteristic of a continuous silicene layer as expected from theoretical calculations [34]. Furthermore, the electrical characterization of this silicene transistor device displays ambipolar electron–hole symmetry as expected from a silicene layer [35]. These experimental and theoretical results are in line with our model of a continuous silicene layer with only topological defects.

## 6. Conclusion

With the help of various STM images recorded at different magnifications, we propose a geometrical model allowing a better understanding of the (2√3x2√3)R30° structures obtained by us and by other groups. This new atomic model of the silicene layer is based on periodic arrangements of perfect areas of (2√3x2√3)R30° surrounded by defect areas on a rigid lattice of silver. In the model, the perfect areas of silicene layers are slightly contracted due to a strain epitaxy, whereas the defect areas are attributed to a local relaxation of this strain. It is important to note that a small expansion of the silicene layer induces a contraction in the STM images, therefore the defect areas appear as shrunken hexagons. Such a model is in line with a strong localized interaction between silver and silicon atoms in agreement with recent theoretical calculations [26,27]. This shows that despite the impression of disorder given by the STM images, the silicene film would be a continuous honeycomb layer with only local and periodic deformations.

Moreover, this model allows a comprehensive interpretation of LEED patterns observed by us and by other groups. A generalization of this model explains the main observations of this structure: deviation of its average direction, Moiré patterns and apparent global disorder. The same approach will be used in the interpretation of STM images for ($\sqrt{13}\times\sqrt{13}$)R13.9° structure in a forthcoming paper [37].

**Acknowledgments**

This work was partially supported within the project SyProSi of the Provence-Alpes-Côte-d'Azur region.